\documentclass[aps,pra,twocolumn,10pt,a4paper]{revtex4-1}

\usepackage{graphicx}
\usepackage{enumerate}
\usepackage{amsmath}

\newcommand{\Uion}[1]{\ensuremath{\textrm{U}^{#1+}}}
\newcommand{\ket}[1]{\ensuremath{\left| #1 \right>}}
\newcommand{\rme}[3]{\ensuremath{\langle#1||#2||#3\rangle}}
\newcommand{\vect}[1]{\mathbf{#1}}
\newcommand{\Gspr}{\ensuremath{\Gamma_\textrm{spr}}}
\newcommand{\Cksq}[2]{\ensuremath{\overline{|C^{(#1)}_{#2}|^2}}}
\newcommand{\BdownTK}{\ensuremath{B(\tau K, m\rightarrow g)}}
\newcommand{\BdownE}{\ensuremath{B(E3, m\rightarrow g)}}
\newcommand{\E}[1]{\ensuremath{\times 10^{#1}}}
\newcommand{\eref}[1]{(\ref{#1})}
\newcommand{\Fig}[1]{Fig.~\ref{#1}}

\def\epl{Europhys. Lett.}
\def\jpb{J. Phys. B: At. Mol. Opt. Phys.}
\def\natphot{Nat. Phot.}
\def\natphys{Nat. Phys.}
\def\sci{Science}
\def\rsi{Rev. Sci. Instrum.}
\def\zpa{Z. Phys. A}
\def\jetplett{JETP Lett.}
\def\njp{New J. Phys.}
\def\plb{Phys. Lett. B}

\begin{document}

\title{Resonant electronic-bridge excitation of the $^{235}$U nuclear transition in ions with chaotic spectra}
\author{J. C. Berengut}
\affiliation{School of Physics, University of New South Wales, NSW 2052, Australia}
\affiliation{Max-Planck-Institut f\"ur Kernphysik, Saupfercheckweg 1, 69117 Heidelberg, Germany}

\begin{abstract}

Electronic bridge excitation of the 76 eV nuclear isomeric state in $^{235}$U is shown to be strongly enhanced in the U$^{7+}$ ion, potentially enabling laser excitation of this nucleus. This is because the electronic spectrum has a very high level density near the nuclear transition energy that ensures the resonance condition is fulfilled.
We present a quantum statistical theory based on many-body quantum chaos to demonstrate that typical values for the electronic factor increase the probability of electronic bridge in $^{235}$U$^{7+}$ by many orders of magnitude. We also extract the nuclear matrix element by considering internal conversion from neutral uranium. The final electronic bridge rate is comparable to the rate of the Yb$^+$ octupole transition currently used in precision spectroscopy.

\end{abstract}

\date{5 December 2018}

\maketitle

\section{Introduction}

Precision laser spectroscopy of nuclear transitions will allow an unprecedented probe of nuclear physics, bridging the fields of nuclear and atomic physics. Proposed applications include nuclear lasing~\cite{tkalya11prl}, nuclear quantum optics~\cite{burvenich06prl}, and extremely accurate nuclear clocks~\cite{peik03epl,campbell12prl}. Recent theoretical~\cite{campbell12prl,tkalya15prc,flambaum06prl,litvinova09prc,berengut09prl,karpeshin17prc,porsev10prl} and experimental~\cite{jeet15prl,vonderwense16nat,thielking18nat} work in this direction has focussed on the $^{229}$Th nucleus, which has the smallest known nuclear transition from the ground state --- expected to be in the vicinity of 7.8~eV~\cite{beck07prl}, although the precise energy is still uncertain.

After $^{229}$Th the next lowest-energy nuclear excitation, and the only other known to lie below 1~keV, is the 76~eV nuclear transition of $^{235}$U. This transition has received far less attention because its energy is in the extreme ultraviolet (EUV) and it is a much weaker ($E3$) transition than the $^{229}$Th ($M1$) transition. However it also has some advantages: its location and properties are quite well known compared to the $^{229}$Th isomeric transition (to $\sim0.5$~eV~\cite{browne14nucl-data-sheets}); $^{235}$U is more readily available than $^{229}$Th; $^{235}$U has a very long half-life; and chemical compounds of uranium are available to, for example, load atomic traps.
It is also worth noting that the $^{235}$U transition involves a change in nuclear shell: the Nilsson quantum numbers of the ground and metastable states are $7/2^-[743]$ and $1/2^+[631]$, respectively. Therefore the uranium EUV transition provides a very different probe of nuclear physics than the $^{229}$Th transition.

The major drawback of $^{235}$U for nuclear spectroscopy is that its frequency is huge by laser standards, and well outside the conventional range. Nevertheless there have been recent demonstrations of up-conversion of frequency combs using high-harmonic generation that can achieve EUV frequencies~\cite{kandula10prl,cingoz12nat,morgenweg14natphys,porat18natphot}.

The other issue is that with a natural transition lifetime of order $10^{24}$~seconds, the $^{235}$U nuclear transition is considered too weak for precision spectroscopy (see, e.g.~\cite{vonderwense16nat}). In this work we show that, by carefully selecting suitable ions and using the electrons to mediate the nuclear transition via electronic bridge (EB), the strength of this nuclear transition can be brought into the range of existing atomic transitions used as frequency standards.

In the electronic bridge process, a nuclear decay occurs not by the direct emission of a photon, but rather by the excitation of an electron, which in turn decays via photoemission.
Despite being a third-order $\gamma$-radiation process in QED (see \Fig{fig:EB_feyn}), the electronic bridge process can be the dominant channel for the decay of a nuclear isomer, particularly if a resonance channel is available~\cite{krutov68ann-der-phys}. This also applies to the inverse process, sometimes called ``inverse electronic bridge''~\cite{tkalya90sov-phys-dokl}. The key point is that the nucleus only weakly couples to low-energy photons due to the small size of the nucleus in comparison to the wavelength of the radiation, while electrons can act as effective mediators of the interaction. EB has previously been studied in $^{235}$U~\cite{hinneburg79zpa,hinneburg81zpa,tkalya90sov-phys-dokl}, $^{229}$Th~\cite{tkalya92jetplett,tkalya92sjnp,kalman94prc,karpeshin99prl,kalman01prc,porsev10pra,porsev10pra0}, and for the 3.4~keV excited-state nuclear transition in $^{84}$Rb~\cite{tkalya14prc}. Laser-induced electronic bridge has been proposed to determine the excitation energy of the $^{229}$Th isomer in~\cite{karpeshin92plb,porsev10prl,bilous18njp}.
Nevertheless, as yet there is no clear experimental observation of the EB mechanism~\cite{tkalya04laser-phys}.

In this manuscript we envisage laser excitation of the $^{235}$U nucleus in a trapped ion via the EB mechanism. In any such attempt it is necessary to suppress further photoionisation by the 76~eV photons (as well as the internal conversion decay mode of the nuclear isomer). Therefore it is necessary to strip $^{235}$U of at least its six valence electrons. 
%This also allows the use of very high laser intensities without risking multiphoton ionization.
The spectral density at 76~eV drops rapidly with increasing ionisation stage. However, in this Letter we show that \Uion{7} should be a very good candidate for nuclear excitation via the EB process because it has a very dense electronic spectrum that ensures the resonance condition is fulfilled. This density is due to having several active $6p$ electrons and relatively low-energy excited orbitals.
Precision spectroscopy of highly charged ions is currently being pursued~\cite{kozlov18arxiv} and the sympathetic cooling of highly charged ions in a cryogenic Paul trap has already been demonstrated~\cite{schwarz12rsi,schmoger15sci}.

\section{Nuclear properties}

The $^{235}$U nuclear ground state has spin and parity $I^P = 7/2^-$, while the low-energy metastable state is $1/2^+$. In order to calculate properties of the $E3$ transition we require the reduced nuclear matrix element $B(E3, m\rightarrow g)$. This can be obtained by considering the internal conversion of the $^{235}$U atom, which has a half-life of approximately 26~min~\cite{browne14nucl-data-sheets}.
Following the conventions of \cite{porsev10pra} we define the hyperfine-interaction Hamiltonian between nuclear operators $M^\lambda_K$ and usual electronic hyperfine-interaction operators $T_{K\lambda}$ as
\begin{equation}
H_\textrm{int} = \sum_{K\lambda} M^\lambda_K T_{K\lambda} .
\end{equation}
The reduced operators of $M$ are related to the usual nuclear matrix elements by
\[
B(\tau K, m\rightarrow g) = \frac{2K+1}{4\pi}
	\frac{\left|\rme{g}{M_K}{m} \right|^2}{2I_m + 1}
\]
where $K$ is the interaction multipolarity. \BdownTK\ is usually measured in Weisskopf units (see, e.g.~\cite{bohr98book1}). For $E3$ transitions the Weisskopf unit is $B(E3) = 0.05940 A^2~e^2\textrm{fm}^6 = 1.494\times10^{-25}$ in atomic units ($\hbar = e = m_e=1$; $A=235$).

In this letter we neglect the hyperfine splitting of levels, therefore the total wavefunction can be factorised into nuclear and electronic parts (this is equivalent to averaging over the hyperfine structure). Internal conversion involves a relaxation of the nucleus ($m\rightarrow g$) with a simultaneous emission of an electron from the shell $\alpha$. For uranium in the ground electronic state $5f^3 6d 7s^2\ ^5L_6^o$, the corresponding internal conversion rate is
\begin{equation}
\label{eq:IC_CA}
\Gamma_\textrm{IC} = \frac{8\pi^2}{[K]^2} B(\tau K, m\rightarrow g)
\sum_\alpha \frac{n_\alpha}{[j_\alpha]} \sum_{jl} 
\left| \rme{\alpha}{T_K}{\varepsilon jl} \right|^2 .
\end{equation}
Here we have introduced the notation $[j] = 2j+1$, $n_\alpha$ is the initial occupancy of the shell $\alpha$, and the emitted electron has energy $\varepsilon$.

A configuration interaction calculation using the atomic code \texttt{AMBiT}~\cite{kahl18arxiv} indicates initial shell occupancies for the uranium ground state of $7s^{1.79}\,6d_{3/2}^{1.09}\,6d_{5/2}^{0.12}\,5f_{5/2}^{2.73}\,5f_{7/2}^{0.27}$. With these values of $n_\alpha$, we calculate the electronic factor from \eref{eq:IC_CA} 
\[
\sum_\alpha \frac{n_\alpha}{[j_\alpha]} \sum_{jl} 
\left| \rme{\alpha}{T_K}{\varepsilon jl} \right|^2 = 1.23\E{6}.
\]
The internal conversion is dominated by the contribution of core $6p$ shells with emission of a $d$-wave electron~\cite{grechukhin76sjnp}, unlike in thorium where internal conversion mainly comes from from the $7s$ shell~\cite{bilous18prc}. Using the measured internal conversion lifetime of 26~min we obtain $\BdownE = 0.036$~W.u., consistent with previous calculations~\cite{grechukhin76sjnp}.

%Using the measured internal conversion lifetime of 26~min we obtain $\BdownE = 0.036$~W.u. We have also performed a more sophisticated calculation of the internal conversion coefficient including configuration interaction for both U and all possible final levels of \Uion{} which gives the same result to within 1\%. Our results are consistent with previous calculations~\cite{grechukhin76sjnp}.

At this point it is worthwhile to make a brief aside and calculate the natural linewidth of the transition. Using the standard formula \cite{bohr98book1,porsev10pra} we obtain
\begin{align}
\Gamma_\gamma (E3, m\rightarrow g) &= \frac{8\pi}{(7!!)^2} \frac{4}{3} \left( \frac{\omega_N}{c} \right)^{7} \BdownE \nonumber \\
&= 1.0\E{-24}~\textrm{s}^{-1} .
\label{eq:naturalwidth}
\end{align}
The natural lifetime is therefore much larger than the half-life of the $^{235}$U nucleus, and is even longer than the lifetime of the Universe. However, this longevity is only realised in special systems, such as a bare uranium nucleus: electronic bridge interactions will generally dominate. Indeed, the mere presence of atomic electrons may induce virtual internal conversion rates several orders of magnitude larger than suggested by \eref{eq:naturalwidth}~\cite{hinneburg79zpa,hinneburg81zpa}.

\section{Electronic spectrum}

In order to overcome the smallness of \eref{eq:naturalwidth}, and enable laser spectroscopy of this nucleus, we seek an electronic structure which maximises the electronic bridge mechanism.
In this work we concentrate on \Uion{7}, which has ground state configuration [Hg]~$6p^5\ ^2P_{3/2}^o$. The lowest excited states are the fine-structure partner $6p^5\ ^2P_{1/2}^o$ and the $6p^4 5f$ levels, which are some 14~eV above the ground state. In order to excite the $E3$ nuclear isomeric transition using EB we require an $E3$ electronic hyperfine transition from the ground state. Therefore we require even-parity levels with $3/2 \leq J \leq 9/2$ in the region of 76~eV.

\begin{figure}[b]
	\includegraphics{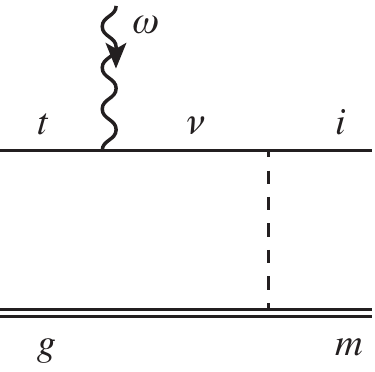}
	\caption{\label{fig:EB_feyn}
		Feynman diagram of the considered electronic bridge process. The double line represents the nucleus, while the dashed line is the hyperfine-$E3$ interation. In our excitation scheme the electronic state $t$ is the lowest $6p^4 5f\ [5/2]^-$ level and $i$ is the $6p^5\ [3/2]^-$ ground state.}
\end{figure}

In our scheme, we would first populate the $6p^4 5f\ J = 5/2^-$ metastable state. This level has only a suppressed $M1$ transition to the ground state (because $\Delta l = 2$ it proceeds only via configuration mixing). This could be populated directly, or via the $6p^46d$ levels at around 27~eV. We would then excite the system with a 62~eV light source to an even parity level $\nu$, which could in turn decay to the ground state with nuclear excitation (see Fig.~\ref{fig:EB_feyn}). This scheme maximises the number of levels that participate in the EB process.

The even-parity spectrum of \Uion{7} begins with the $6p^46d$ configurations and rapidly becomes very dense with increasing energy. At 76~eV above the ground state, the density is over 2000 levels per eV, or $\sim 30$ per eV for each subspace with even parity and fixed angular momentum and projection. At this energy the average mixing between states (i.e. the root-mean-square off-diagonal Hamiltonian matrix element $(\overline{H_{ij}^2})^{1/2}$~\cite{flambaum94pra})
is around 5 times larger than the level spacing, which means that the levels are essentially completely mixed. Under these conditions we have many-body quantum chaos (MBQC), and a statistical description of the system becomes valid~(see, e.g.~\cite{flambaum93prl,flambaum94pra,gribakin99austjphys,gribakin03jpb,dzuba12pra1,dzuba13pra} and references within).
MBQC in electronic spectra has previously been predicted in near-neutral lanthanides~\cite{flambaum94pra} and actinides~\cite{dzuba10prl}, as well as at high excitation energies in highly charged ions with open $f$-shells~\cite{flambaum02pra,gribakin03jpb}.

In the quantum statistical theory we express the chaotic even levels $\ket{\nu}$ in the basis of principal components $\ket{k}$ as
\begin{equation}
\label{eq:nu_to_k}
\ket{\nu} = \sum_k C^{(\nu)}_k \ket{k}
\end{equation}
where the coefficients $C^{(\nu)}_k$ behave as uncorrelated random variables with mean zero ($\overline{C^{(\nu)}_k} = 0$) and 
\begin{gather}
\overline{C^{(\nu)}_k C^{(\mu)}_m} = \delta_{\nu\mu}\delta_{km} \overline{|C^{(\nu)}_k|^2} \\
\label{eq:C_k^2}
\Cksq{\nu}{k} = \frac{D_J}{2\pi}\frac{\Gspr}{(\epsilon_\nu - \epsilon_k)^2 + \Gspr^2/4}
\end{gather}
where $\Gspr$ is known as the spreading width, which depends only weakly on energy~\cite{gribakin03jpb}.

In a `configuration-averaged' statistical theory the principal components \ket{k} can be configurations. However in order to preserve the exact angular properties of the levels and operators, in this Letter we use functions with definite values of $J$ and projection $M$ built by performing a configuration interaction calculation using all configuration state functions belonging to a single non-relativistic configuration. Previously we used this ``level resolved'' statistical theory to calculate electron-capture cross-sections in W$^{20+}$~\cite{berengut15pra}.

\section{Electronic bridge}

Again, we neglect the hyperfine splitting of levels and factorise the nuclear and electronic parts of the $^{235}$\Uion{7} wavefunction.
Following the notation of~\cite{porsev10prl,porsev10pra0} we can write the rate of the spontaneous EB process as
\begin{align}
\Gamma_\textrm{EB} &= \frac{4\omega^3}{3c^3} \frac{\left| \rme{I_g}{M_K}{I_m} \right|^2}{[K][I_m][J_t]} G_2 \nonumber \\
&= \frac{16\pi}{3 [K]^2[J_t]} \frac{\omega^3}{c^3}\BdownE\,G_2
\label{eq:Gamma_EB}
\end{align}
where $\omega$ is the frequency of the absorbed photon.
%where $[J] \equiv 2J+1$ and $K=3$ for the $^{235}$U nuclear isomeric transition. 
The electronic factor is
\begin{equation}
\label{eq:G2_nu}
G_2 = \sum_{J_\nu} \frac{1}{[J_\nu]} \left| \sum_\nu \frac{\rme{i}{T_3}{\nu}\rme{\nu}{\vect{d}}{t}}{\omega_{\nu i}-\omega_N + i\Gamma_\nu/2} \right|^2
\end{equation}
where $\omega_{\nu i} = \epsilon_\nu - \epsilon_i$ and $\omega_N = E_m - E_g \approx 76$~eV. $T_3$ is the rank-3 electronic hyperfine interaction operator (see, e.g. Appendix B of \cite{beloy08pra}), and $\vect{d}$ is the electric dipole operator.

We now apply the statistical theory of MBQC to the EB process, substituting Eqs.~\eref{eq:nu_to_k} -- \eref{eq:C_k^2} into \eref{eq:G2_nu}. We obtain three terms which, following the nomenclature created for atomic processes in~\cite{flambaum15pra}, we call the coherent, independent resonance, and residual, respectively:
\begin{align}
\label{eq:G2_coh}
G_2 = &\sum_{J_\nu} \frac{1}{[J_\nu]} \left(\ 
\left| \sum_{\nu k} \Cksq{\nu}{k} \frac{\rme{i}{T_3}{k}\rme{k}{\vect{d}}{t}}{\omega_{\nu i}-\omega_N + i\Gamma_\nu/2} \right|^2 \right. \\
\label{eq:G2_IR}
& + \sum_{\nu k m} \Cksq{\nu}{k}\, \Cksq{\nu}{m}
\frac{|\rme{i}{T_3}{k}|^2|\rme{m}{\vect{d}}{t}|^2}{(\omega_{\nu i}-\omega_N)^2 + \Gamma_\nu^2/4} \\
\label{eq:G2_res}
& + \left. \sum_\nu \left| \sum_k \Cksq{\nu}{k}
\frac{\rme{i}{T_3}{k}\rme{t}{\vect{d}}{k}}{\omega_{\nu i}-\omega_N + i\Gamma_\nu/2}
\right|^2\ \right) .
\end{align}
In our case the independent-resonance (IR) contribution \eref{eq:G2_IR} is larger than the coherent and residual by two orders-of-magnitude, therefore we neglect the latter.

Expanding the $\Cksq{\nu}{k}$, we obtain expressions that contain sums over $\nu$ which only manifest in the energy denominators, $\omega_{\nu i}$. Because of the energy conservation condition, the EB width is much smaller than the mean level spacing. Therefore the EB process will be dominated by only a few resonances near $\omega_N$. We may estimate a typical ``unlucky'' case where $\omega_N$ lands exactly between two levels amongst a forest of levels separated by $D_J$. Then
\[
\sum_\nu \frac{1}{(\epsilon_{\nu i} - \epsilon_N)^2} 
\approx \sum_n \frac{1}{(n+1/2)^2D_J^2} = \frac{\pi^2}{D_J^2}
\]
and we obtain for the independent-resonance contribution
\begin{multline}
\label{eq:G2_final}
G_2^\textrm{IR} = \sum_{J_\nu = 3/2}^{7/2} \frac{1}{[J_\nu]}\frac{1}{4}
\sum_k \frac{|\rme{i}{T_3}{k}|^2\,\Gspr}{(\omega_{ki} - \omega_N)^2 + \Gspr^2/4} \\
\times\sum_{m}\frac{|\rme{m}{\vect{d}}{t}|^2\,\Gspr}{(\omega_{mi} - \omega_N)^2 + \Gspr^2/4}.
\end{multline}
Note that this procedure is different to that presented in~\cite{flambaum15pra} for the calculation of atomic processes such as photoexcitation and photoionization. In that work the process of averaging over a photon energy with width $\Delta\omega$ containing a large number of resonances allowed the authors to replace the summation over $\nu$ with an integral over energy. In that case one obtains a prefactor $\sim D_J/\Gamma_{\nu}$ in the IR and residual terms, which is not present in our very narrow EB process.

\section{Results and Discussion}

We have calculated \eref{eq:G2_final} using \texttt{AMBiT}. Core orbitals were calculated by solving the self-consistent Dirac-Hartree-Fock equations in the $V^N$ approximation, including core electrons up to $6s^2 6p^5$. Excited orbitals were generated in the $V^{N-1}$ potential of the residue $6s^2 6p^4$, and then orthogonalised to the core orbitals using a Gram-Schmidt procedure. Orbitals with principal quantum numbers up to 10 and $l \leq 4$ ($g$-wave) were included in the calculation.

Principal components \ket{k} were generated as follows. First, we generate configuration state functions (CSFs) from all possible configurations with configuration-averaged energy below 128~eV from the ground state. We then diagonalise Hamiltonian submatrices consisting of all CSFs belonging to a single non-relativistic configuration. The resulting eigenstates are our \ket{k}. These states still preserve the angular momentum and projection from the CSFs, but are more realistically distributed in energy space because they are spread out by the configuration mixing~\cite{berengut15pra}.

To determine the spreading width $\Gspr$ and level density $D_J$ we created Hamiltonian matrices for $J^\pi = 3/2^+$, $5/2^+$, and $7/2^+$ including all principal components. We find $\Gspr = 2\pi \overline{H_{ij}^2}/D_J \approx 4.8$~eV, where $H_{ij}$ is the Hamiltonian matrix element~\cite{flambaum94pra}.
%We estimated $\Gspr$ in two ways. First, we used the formula $\Gspr = 2\pi \overline{H_{ij}^2}/D_J \approx 4.8$~eV, where $H_{ij}$ is the Hamiltonian matrix element. In the second method we diagonalise the Hamiltonian to obtain values of $C^{(\nu)}_{k}$, and then estimate $\Gspr$ from their distribution~\cite{flambaum94pra}. The second method gives results that fluctuate somewhat because they are sensitive to binning parameters, but they are consistent to within $\sim 20\%$ with the results of the first method.

Using these principal components and $\Gspr$, our calculation of \eref{eq:G2_final} yields $G_2 = 8.0\times10^6$. This value is not sensitive to the exact values of $\omega_N$ and $\Gspr$ since each term in \eref{eq:G2_final} integrates over a dense set of principal components within $\Gspr$ of $\omega_N$.

\begin{figure}[tb]
\includegraphics[width=0.45\textwidth]{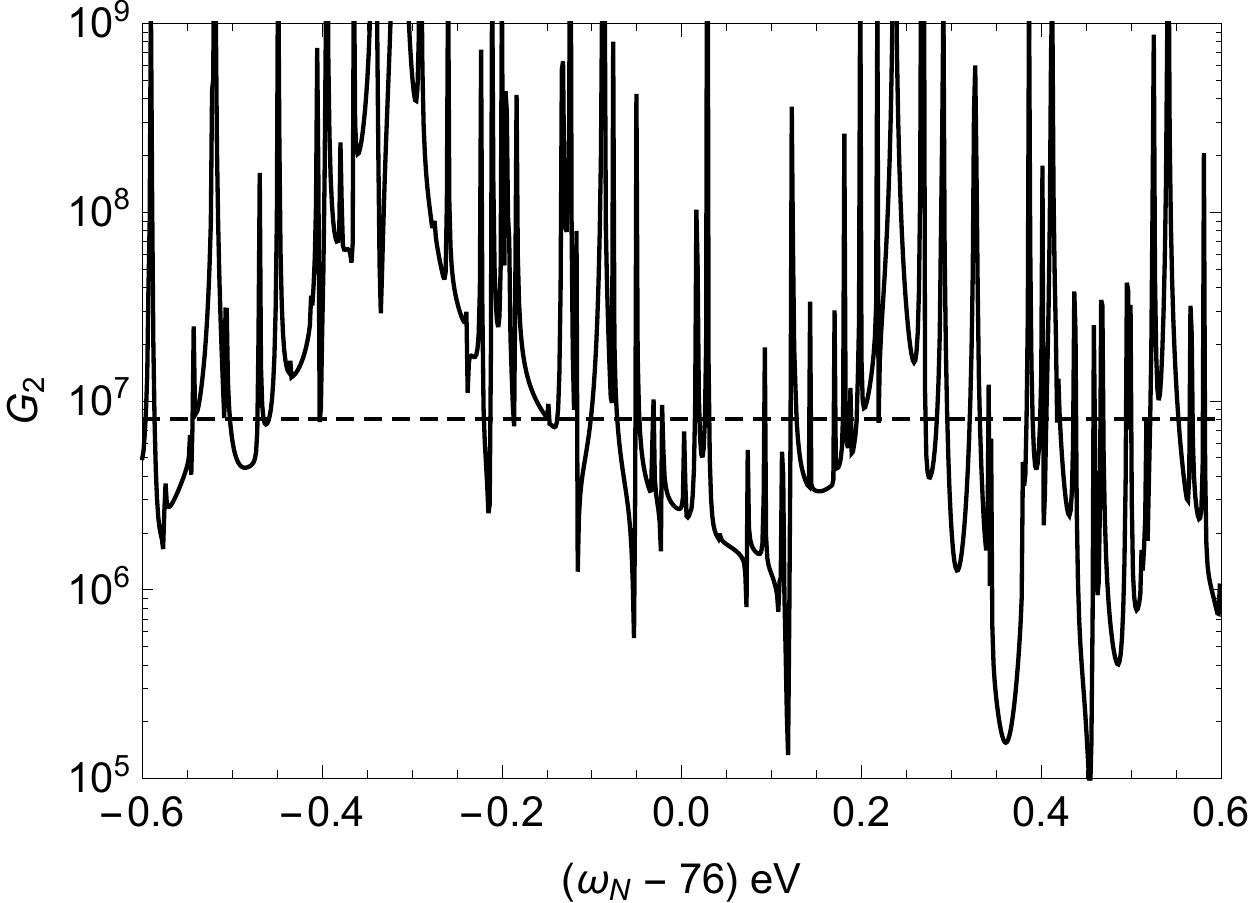}
\caption{\label{fig:G2_exact} $G_2$ calculated using \eref{eq:G2_nu} and a configuration interaction spectrum generated using \texttt{AMBiT}. For comparison, the value calculated with the statistical theory of many-body quantum chaos~\eref{eq:G2_final} is shown $G_2 = 8.0\times10^6$ (dashed line).
}
\end{figure}

\begin{figure}[tb]
\includegraphics[width=0.45\textwidth]{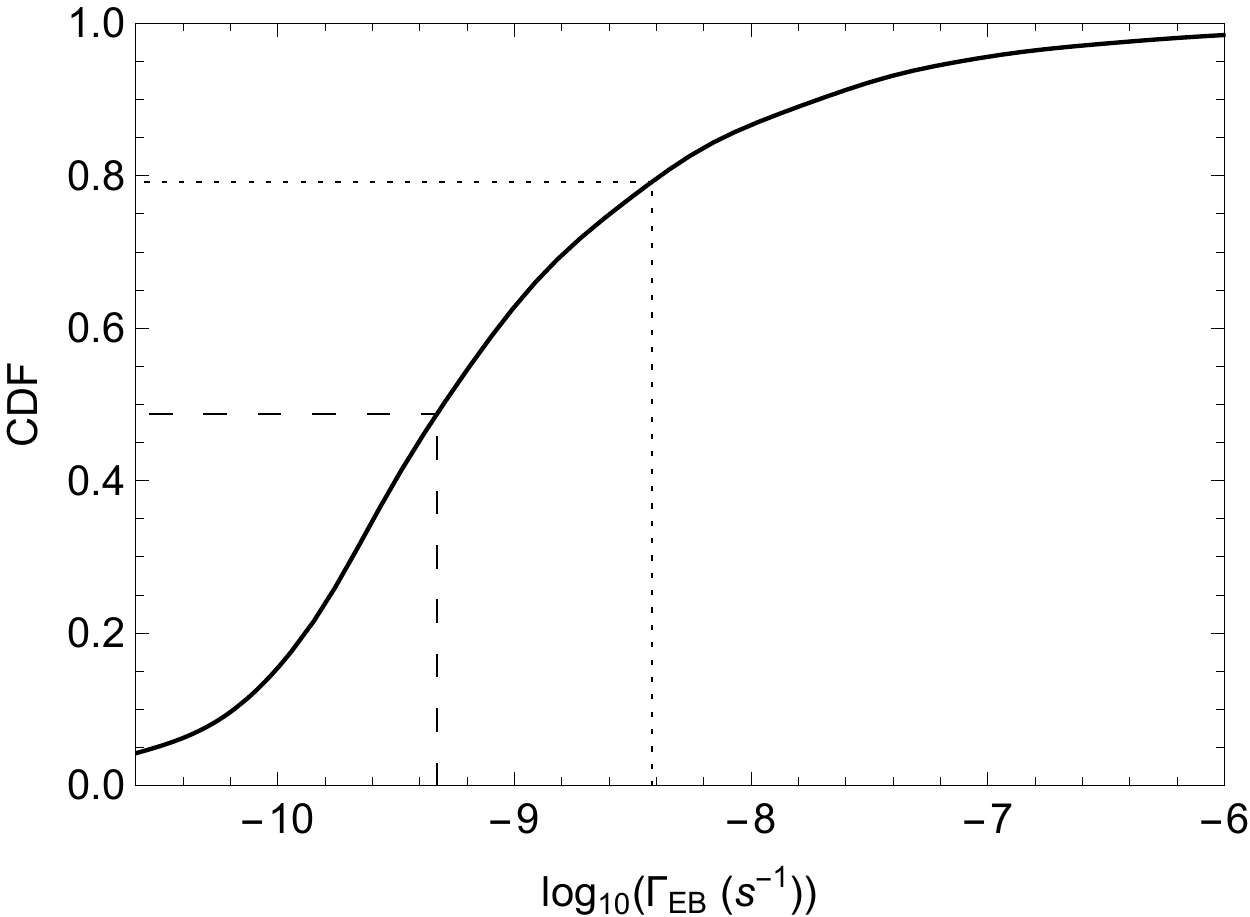}
\caption{\label{fig:GammaEB} Cumulative distribution function for $\Gamma_\textrm{EB}$ extracted from \Fig{fig:G2_exact}. Dashed line: estimate of $\Gamma_\textrm{EB}$ from MBQC \eref{eq:GammaEB_stat}; Dotted line: linewidth of $E3$ clock transition in Yb$^+$~\cite{biemont98prl}. }
\end{figure}

To check our statistical theory, we have used \texttt{AMBiT} to generate a ``complete'' calculation of the even levels $\nu$ near $\omega_N$ using configuration interaction (exact diagonalization of the Hamiltonian matrix). Due to MBQC, the resulting eigenstates $\nu$ and energies $\epsilon_\nu$ only represent the real spectrum in a statistical sense. That is, the generated matrix $H$ is an instance of the random matrix with correct average spacing and mixing, and the resulting spectral components are only a single instance of the random variables $C^{(\nu)}_k$. Using the spectrum thus obtained in \eref{eq:G2_nu} we generated $G_2$ as a function of $\omega_N$ in the vicinity of 76~eV. 
The results are shown in~\Fig{fig:G2_exact}. The positions of resonances and their strengths are only indicative; nevertheless the MBQC calculation falls near the median value of $G_2$ (see \Fig{fig:GammaEB}) supporting the validity of the statistical approach.


With our calculated values of \BdownE\ and $G_2$, we are now able to estimate the total electronic bridge rate of \Uion{7} from \eref{eq:Gamma_EB}:
\begin{equation}
\label{eq:GammaEB_stat}
\Gamma_\textrm{EB} \approx 4.7 \E{-10}\ \textrm{s}^{-1} .
\end{equation}
This rate is comparable to that of narrow atomic transitions used in precision spectroscopy, for example the atomic $E3$ transition of Yb$^+$~\cite{godun14prl,huntemann16prl}.
Of course, since we do not know the precise positions of either the nuclear transition or the electronic resonances, the real value of $G_2$ (and hence $\Gamma_\textrm{EB}$) may be orders of magnitude larger. To quantify this uncertainty, in \Fig{fig:GammaEB} we present a cumulative distribution function for $\Gamma_\textrm{EB}$ based on the values of \Fig{fig:G2_exact}.

\section{Conclusion}

We have shown that by careful selection of ion stage and electronic bridge scheme, the effective strength of the nuclear transition in $^{235}$U can be increased by many orders of magnitude. This brings the transition width to within the range of current atomic experiments. Different ion stages will allow the electronic bridge to be adjusted further, depending on how close the nuclear transition is to an electronic resonance.
Many-body quantum chaos is also be present at 76~eV in \Uion{8}, and this may be useful if \Uion{7} is not favorable (for example, if the nuclear resonance falls far from a suitable electronic level, suppressing $G_2$).
Other charge stages may also allow for useful interplay between electrons and nuclei.

\acknowledgments

I thank the following people for very useful discussions: P. Bilous, A. P\'alffy, E. Peik, L. von der Wense, C. Schneider, P. Thirolf, P. O. Schmidt, V. Flambaum, O. Versolato, and J. Crespo L\'opez-Urrutia.
This work was supported by the Alexander von Humboldt Foundation.

\bibliography{references}

\end{document}